\documentclass[journal]{IEEEtran}
\ifCLASSINFOpdf
\else
\fi
\usepackage{amssymb} 

\usepackage[cmex10]{amsmath}
\usepackage{algorithmic}

\usepackage{array}
\usepackage{url}

\usepackage{comment}

\ifCLASSOPTIONcompsoc
\usepackage[caption=false,font=normalsize,labelfon
t=sf,textfont=sf]{subfig} \else
\usepackage[caption=false,font=footnotesize]{subfig} \fi
\usepackage{mdwmath}
\usepackage{mdwtab}
\usepackage{multirow}
\usepackage[ruled,vlined]{algorithm2e}
\usepackage{graphicx}
\usepackage{epstopdf}
\usepackage[justification=centering]{caption}
\usepackage[section]{placeins}
\usepackage{float}

\usepackage{xcolor}

\begin{document}

%
\title{Tango: A Deep Neural Network Benchmark Suite for Various Accelerators}

\author{\IEEEauthorblockN{Aajna Karki, Chethan Palangotu Keshava, Spoorthi Mysore Shivakumar, \\ Joshua Skow, Goutam Madhukeshwar Hegde, Hyeran Jeon} \\
 \IEEEauthorblockA{Computer Engineering Department\\
 San Jos\'{e} State University\\
 San Jos\'{e}, CA, USA \\
 Email: \{aajna.karki, chethan.keshava, spoorthi.mysoreshivakumar, \\joshua.skow,  goutammadhukeshwar.hegde,  hyeran.jeon\}@sjsu.edu}
 }

\maketitle
    
\begin{abstract}
Deep neural networks (DNNs) have been proving the effectiveness in various computing fields. To provide more efficient computing platforms for DNN applications, it is essential to have evaluation environments that include assorted benchmark workloads. Though a few DNN benchmark suites have been recently released, most of them require to install proprietary DNN libraries or resource-intensive DNN frameworks, which are hard to run on resource-limited mobile platforms or architecture simulators. To provide a more scalable evaluation environment, we propose a new DNN benchmark suite that can run on any platform that supports CUDA and OpenCL. The proposed benchmark suite includes the most widely used five convolution neural networks and two recurrent neural networks. We provide in-depth architectural statistics of these networks while running them on an architecture simulator, a server- and a mobile-GPU, and a mobile FPGA. 
\end{abstract}


\IEEEpeerreviewmaketitle

\section{Introduction}\label{sec:introduction}
Neural network has regained huge attention from industry and academia recently. Neural-network-based applications enable machines to automatically recognize individuals in photos, cars to navigate by themselves, and medical devices to diagnose cancer. Thanks to the birth of massively-parallel computing platforms such as GPUs and specialized hardware accelerators, complex cognitive applications powered by neural networks provide real-time responses with high prediction accuracy. While application and algorithm researchers are leading the advancement of neural networks for better prediction accuracy, computer architects and hardware designers are trying to optimize the computing platforms for providing more efficient neural network computing environment. 

To optimize hardware platforms, it is essential to have an evaluation environment that helps measure architectural characteristics of the neural network models and verify the performance impact of the optimization. For this purpose, computer architecture studies use benchmark suites. For example, SPEC~\cite{spec} and PARSEC~\cite{parsec} benchmark suites have been used for testing the efficiency of CPU designs by providing an assorted set of widely executed CPU applications. Recently, a few benchmark suites have been developed for deep neural networks (DNN)~\cite{deep_bench, fathom, tbd, dawnbench, tensorflow_bench, mlperf}. However, most of them require installation of DNN frameworks (i.e. TensorFlow~\cite{tensorflow} or Keras~\cite{keras}) and proprietary DNN libraries (i.e. NVIDIA cuDNN~\cite{cudnn}), which are hard to be deployed on platforms and architecture simulators that do not support the libraries due to insufficient system resources or compatibility issue. Therefore, there is a high demand for a new benchmark suite that provides widely used DNN workloads and can run without proprietary libraries or heavy frameworks. 

In this paper, we present a new DNN benchmark suite which does not require DNN framework or proprietary library installation. The benchmark suite provides a set of most widely used convolutional neural networks (CNNs) and recurrent neural networks (RNNs) written in CUDA C and OpenCL. With this benchmark suite, accelerator designers and mobile platform researchers can test their new architectures with various CNN and RNN models, which was hard or impossible due to the requirement of libraries or DNN frameworks. This new benchmark suite also can be used for testing a new accelerator at early design phase before developing an interface for DNN frameworks, which will help improve the productivity. 

This paper makes the following contributions:
\begin{itemize}
\item To our best knowledge, this is the very first DNN benchmark suite that is purely written in CUDA C and OpenCL. The benchmark suite provides code for inference phase of five CNNs (CifarNet~\cite{cifarnet}, AlexNet~\cite{alexnet}, SqueezeNet~\cite{squeezenet}, ResNet~\cite{resnet}, and VGGNet~\cite{vggnet}) and two RNNs (GRU~\cite{gru} and LSTM~\cite{lstm}) by decomposing neural network operations into fundamental mathematical computations.

\item This new benchmark suite supports all architectures or software simulators that can run CUDA and OpenCL applications. Thus, we believe this benchmark suite can be widely used by computer architects as well as neural network algorithm researchers for evaluating their ideas in both hardware and software levels. 

\item We provide in-depth architectural characteristics of individual networks by running them on a GPU architecture simulator (GPGPU-Sim~\cite{gpgpusim}), a server GPU (NVIDIA GK210~\cite{gk210}), a mobile GPU (NVIDIA TX1~\cite{tx1}), and an FPGA (Xilinx PynQ-Z1~\cite{pynq}). Unlike previous studies that mainly relied on vendor-provided system profilers, we were able to examine networks with various configurations, such as different cache size and schedulers, thanks to the compatibility with the architecture simulator. The provided statistics will provide insights to the researchers.

\end{itemize}

\section{Motivation and Related Work}\label{sec:motivation}

\subsection{General-Purpose GPU Benchmark Suites}
Since GPU has been used for general-purpose computing, a few GPU benchmark suites have been developed to evaluate performance and architectural characteristics of GPUs for general purpose applications. NVIDIA has been releasing CUDA SDKs~\cite{cuda_sdk} when they release a new CUDA version. NVIDIA CUDA SDK includes various scientific and graphics application kernels as well as a few sample device interfacing applications that help CUDA programmers learn new features of CUDA. Rodinia benchmark suite~\cite{rodinia} was developed by a team of University of Virginia. It includes CUDA and OpenCL kernels of graph, data mining, image recognition, bioinformatics, and physics simulation domains. They also provide CPU version kernel code that are parallelized with OpenMP for a comparison. Parboil~\cite{parboil} was developed by a team of University of Illinois at Urbana-Champaign. Parboil includes CUDA and OpenCL kernels of image processing, biomolecular simulation, fluid dynamics, and astronomy domains. Parboil also includes CPU equivalents of the kernels that are parallelized by OpenMP. GPGPU-Sim~\cite{gpgpusim}, which is a GPU architecture simulator, also encloses an assorted CUDA programming samples. The samples include various scientific and graphics applications such as AES cryptography, graph processing, laplace computation, n-queens solver, and ray tracing. A team of University of Southern California compiled various graph processing kernels written in CUDA~\cite{xu_iiswc}. This graph benchmark suite provides 12 graph-based processing kernels including graph coloring, graph cuts, graph clustering, all pairs and single-source shortest path, and page rank, which help explore various characteristics of graph-based applications.

These benchmark suites provide a wide range of general-purpose GPU kernels. However, none of them implements state-of-the-art DNN kernels. GPGPU-Sim benchmark includes \textit{NN} which is a 4-layer DNN for handwriting recognition. Though this application is a good example of pure CUDA-based DNN implementation, the network structure is very simple and hardly represents the state-of-the-art DNNs. Thus, we need a new benchmark suite that reflects widely used state-of-the-art DNN structures to retrieve more realistic insights of DNN execution on GPUs. 
\vspace{-10pt}
\subsection{DNN Benchmark Suites}
As DNN is proved as a powerful solution for various cognitive and characterization problems, a few DNN benchmark suites have been recently released. DNNMark~\cite{dnnmark} provides per-layer computation primitives for DNNs. The supported primitives include forward and backward computations of convolution, pooling, activation, and fully-connected layers. Baidu DeepBench~\cite{deep_bench} includes CNN and RNN core kernels for heterogeneous architectures such as mobile and server CPUs, and NVIDIA and AMD GPUs. The kernels internally call DNN library functions provided by the platform vendors. Thus, the performance of various architectures with the optimized library functions can be compared. Fathom~\cite{fathom} provides TensorFlow scripts for eight DNN models including convolutional, recurrent, and fully-connected neural networks. Based on the fact that most of the DNN frameworks internally use similar DNN libraries such as cuDNN and hence the performance of the same network should be similar across frameworks, they implemented various DNN workloads with one of the most widely used frameworks, TensorFlow. TBD~\cite{tbd} focuses on DNN training. The provided kernels can run on three DNN frameworks, TensorFlow, MXNet, and CNTK, across different hardware configurations (single GPU, multiple GPUs, and multiple machines). It also incorporates an analysis toolchain for resource and performance profiling of these models. DAWNBench~\cite{dawnbench} is a benchmark suite that can test end-to-end performance evaluation for training and inference while tweaking various hyper-parameters. The provided code is written for PyTorch~\cite{pytorch} and TensorFlow. TensorFlow also provides an assorted set of image classification models~\cite{tensorflow_bench}. The TensorFlow script files for five DNN models can run on NVIDIA GPUs. 

These DNN benchmark suites are useful for testing core and full computations of various DNN structures. However, all of them require installation of specialized libraries such as cuDNN and cuBLAS~\cite{cublas} and DNN frameworks. NVIDIA GPUs can run the libraries but architecture simulators and non-CUDA architectures do not support these libraries. Especially, the architecture simulators will hardly support these libraries in the near future because the libraries are proprietary close-source software. Furthermore, many DNN frameworks have high resource demands. For example, it is tricky to install Caffe2~\cite{caffe2} and TensorFlow on resource-limited mobile accelerators such as NVIDIA TX1~\cite{tx1} or Xilinx PynQ~\cite{pynq} due to insufficient memory space. Though the frameworks are getting more compact (i.e. TensorFlow Lite~\cite{tensorflow_lite} supports Raspberry Pi~\cite{raspberry}), many of them are not easily deployable on small accelerators. Thus, computer architecture community cannot use these benchmark suites for evaluating their new architecture design ideas. Therefore, we need a new DNN benchmark suite that does not require any proprietary libraries or resource-hungry DNN frameworks. 
\vspace{-10pt}
\subsection{Why Do We Need Another Benchmark Suite?}
In this paper, we introduce a new DNN benchmark suite for GPUs. This new benchmark suite is written in CUDA C and OpenCL without using DNN libraries. Therefore, the benchmark suite can run on any platform that supports either of these two languages such as GPUs, architecture simulators, and FPGAs. Computer architecture researchers can use this benchmark suite for evaluating their ideas of new accelerator design. Also, DNN algorithm researchers can use this benchmark suite to evaluate new algorithms by simply replacing the core functions of individual layers. Note that for those benchmark suites that use library functions, it is not easy to modify core functions. Currently, our proposed benchmark suite provides feed-forward code for inference phase only but we plan to extend the suite to also provide back-propagation code for training phase. By running both training and inference code on various platforms and simulators, researchers will be able to acquire useful insights.

\section{Neural Networks in the Benchmark Suite}\label{sec:Architecture}

To provide representative DNN workloads, we first identified the most widely used DNNs. Among many deep learning structures, CNN and RNN have been extensively applied for many applications. CNN is mainly used for the applications that need to extract patterns from image inputs such as face recognition and obstacle detection. On the other hand, RNN extracts information from time-series inputs, such as next word prediction and stock price forecasting. To provide high impact workloads, we developed CNNs and RNNs.

Though the effectiveness of CNN and RNN has been well proved, it is not easy to find the optimum network structure for individual applications because the prediction accuracy varies significantly with regarding to the number of layers, size of each layer, activation functions, and many other hyper-parameters. Thus, recent trend is to use one of the well-proved reference models as a baseline and fine-tune the structure to achieve a better accuracy for the target application. Therefore, we implemented the most widely used reference models in this benchmark suite. The proposed benchmark suite currently include CifarNet~\cite{cifarnet, cifarnet2}, AlexNet~\cite{alexnet}, SqueezeNet~\cite{squeezenet}, ResNet~\cite{resnet}, and VGGNet~\cite{vggnet} as CNNs and GRU~\cite{gru} and LSTM~\cite{lstm} as RNNs. All of the seven networks are implemented in CUDA C and CifarNet and AlexNet are also implemented in OpenCL. We are currently developing more networks such as MobileNet~\cite{mobilenet}. Thus, the coverage will keep increasing. 

The selected networks are widely used thanks to the proved performance (many of them are winners of ImageNet Large Scale Visual Recognition Challenge (ILSVRC)~\cite{ilsvrc}). Thus, the existing DNN benchmark suites~\cite{dnnmark, deep_bench, dawnbench, tbd, tensorflow_bench, fathom, mlperf} commonly provide the model files of many of these networks. The networks use different number of layers, number of neurons per layer, activation functions, and types and order of layers. For example, AlexNet uses five convolution layers and three fully-connected layers while ResNet uses 49 convolution layers and one fully-connected layer with shortcut residual paths. Thus, the architectural behaviors of these networks should be different. To evaluate both intra-layer and inter-layer characteristics of individual networks, we implemented the entire structure of the target networks in CUDA C and OpenCL. We used pre-trained model weight files as input of individual layers. The details are discussed shortly.

\subsection{Convolutional Neural Network (CNN)}
CNN is one type of DNN that is used especially for image recognition. 
By adopting the convolution operation that was originally used by image processing for applying image filters, CNN takes an input image and runs convolution operations to extract features. The layers that run convolution operations are called as \textit{convolution layers}. In a convolution layer, a neuron takes a subset of pixel values, multiplies them with a small weight matrix, and sums the weight-scaled values to generate one convolution output. Then, a \textit{pooling layer} follows where the maximum (or average) value of neighboring neurons' convolution results is taken as one final feature value of that region. As only the maximum (or average) value of a few neighboring convolution outputs is taken as a final output, an object can be recognized regardless the position of the object within the input image. Thus, CNN is good for image recognition. Once all features are extracted by a set of convolution layers and pooling layers, one or more \textit{fully-connected layers} are executed to calculate the probability of classification. The combination of convolution layers, pooling layers, and fully-connected layers is the main structure of feed-forward path of a CNN. Back-propagation path consists of layers that compute gradient of prediction errors of forward path execution. 

\subsubsection{\textbf{CifarNet}}
CifarNet~\cite{cifarnet} is developed to recognize objects over CIFAR-10 and CIFAR-100 database~\cite{cifar_db}. It consists of three convolutional layers and two fully connected layers. This network receives three-channel $32\times32$ images as inputs. The last layer consists of as many output neurons as the number of object classes that the model aims to recognize. We used a CifarNet model that is trained for traffic signal detection. The model recognizes nine traffic signals. Thus, the CifarNet code has nine output neurons. This is next fed to a softmax layer to generate the output in terms of probabilities. More details about the pre-trained models can be found in Section~\ref{sec.input}.

\subsubsection{\textbf{AlexNet}}
AlexNet~\cite{alexnet} is the first CNN that proved the efficiency and accuracy of CNN-based object recognition. After the successful debut of AlexNet at the ILSVRC~\cite{ilsvrc} in 2012, extensive studies have been conducted on CNN-based solutions and CNN structure itself. AlexNet consists of five convolutional layers and three fully-connected layers. This network receives three-channel $227\times227$ images as inputs. As our pre-trained model uses ImageNet database~\cite{imagenet} which can recognize 1000 objects, AlexNet consists of 1000 output neurons.

\subsubsection{\textbf{ResNet}}
ResNet~\cite{resnet} or Residual Network was developed by Microsoft and demonstrated a surpassed recognition performance than human on ImageNet database. ResNet has various versions with different number of layers. We developed ResNet-50 that uses 50 layers. ResNet-50 receives three-channel $224\times224$ input images and derives 1000 objects classification. As name says, ResNet uses residuals for training. By incorporating shorcut paths that simply add residuals rather than using the original approximation function result, the network can make the back-propagation to skip a few useless layers. This way, it solves the vanishing problem of DNN.

\subsubsection{\textbf{SqueezeNet}}
SqueezeNet~\cite{squeezenet} is designed to achieve a good accuracy with fewer parameters so that the model can be deployed on embedded platforms. To reduce the model size without sacrificing accuracy, SqueezeNet defines $fire module$ that consists of a $squeeze$ convolution layer that has only $1\times1$ filters and an $expand$ layer that has a mix of $1\times1$ and $3\times3$ convolution filters. With the fire modules, SqueezeNet demonstrated 50 times fewer parameters than the original network~\cite{squeezenet}. SqueezeNet consists of two convolutional layers, eight fire layers and one global pooling layer. The inputs to this network are 3-channel $227\times227$ images. It uses 1000 output neurons.

\subsubsection{\textbf{VGGNet}}
VGGNet~\cite{vggnet} debuted successfully as a winner of ILSVRC~\cite{ilsvrc} in 2014. VGGNet achieved a high accuracy with a very deep network that uses 16 or 19 layers, where each layer uses very small (3 $\times$ 3) convolution filters. We implemented 16-layer VGGNet, which consists of 13 convolution layers, three fully-connected layers, five pooling layers, and one soft-max layer. This network receives three-channel $224\times224$ images as inputs. As our pre-trained model uses ImageNet database~\cite{imagenet} which can recognize 1000 objects, VGGNet consists of 1000 output neurons.

\vspace{-10pt}

\begin{table}[]
\centering
\scalebox{0.8}
{
\begin{tabular}{llll}
\hline
                                                         & \textbf{Input Data}                                                                                            & \textbf{Pre-trained Model}                                                                                                                                     & \textbf{Output}                                                                                                       \\ \hline
\begin{tabular}[c]{@{}l@{}}GRU \& \\ LSTM\end{tabular} & \begin{tabular}[c]{@{}l@{}}Bitcoin stock \\ price values \\ of past two \\ days (scaled)\end{tabular} & \begin{tabular}[c]{@{}l@{}}Trained with bitcoin\\ stock price database\\ (https://www.kaggle.\\ com/team-ai/bitcoin-\\ price-prediction)\end{tabular} & \begin{tabular}[c]{@{}l@{}}Projected next stock \\ price based on past \\ two days' stock price\end{tabular} \\ \hline
CifarNet                                                 & \begin{tabular}[c]{@{}l@{}}Speed limit 35 \\ image\end{tabular}                                       & \begin{tabular}[c]{@{}l@{}}https://github.com/\\ chethankeshava/\\ DeepLearningProject\end{tabular}                                                   & \begin{tabular}[c]{@{}l@{}}Confidence level for \\ all 9 classes\end{tabular}                                \\ \hline
AlexNet                                                  & Cat image                                                                                             & \begin{tabular}[c]{@{}l@{}}https://github.com/BVLC/\\ caffe/tree/master/models/\\ bvlc\_alexnet\end{tabular}                                          & Recognized class id                                                                                          \\ \hline
SqueezeNet                                               & Cat image                                                                                             & \begin{tabular}[c]{@{}l@{}}https://github.com/\\ DeepScale/\\ SqueezeNet/tree/master/\\ SqueezeNet\_v1.0\end{tabular}                                 & Recognized class id                                                                                          \\ \hline
ResNet                                                   & Cat image                                                                                             & \begin{tabular}[c]{@{}l@{}}https://github.com/\\ KaimingHe/\\deep-residual-networks\end{tabular}                                                               & Recognized class id                                                                                          \\ \hline
VGGNet                                                   & Killer whale image
       & \begin{tabular}[c]{@{}l@{}}http://www.robots.ox.ac.uk/\\ $\sim$vgg/research/very\_deep/\end{tabular}
    & Recognized class id                                                                                           \\ \hline
\end{tabular}
}\caption{Input/Output and Pre-trained Models used by networks } \label{tab.input}\vspace{-15pt}
\end{table}


\subsection{Recurrent Neural Network (RNN)}
RNN is another type of DNN that produces an output result that not only depends on the current input but also on a history of previous inputs. RNN has a unique ability of making decisions based on the past incidents, which is similar to a human behavior. Thus RNNs are gaining popularity these days especially for speech or video processing that makes use of sequential information available at the input. RNNs are called recurrent because they perform the same tasks for every element of the sequence. RNN has multiple layers that are stacked in a daisy-chain fashion with the output of one layer given as input for the next layer and each layer repeating the same process. To process such time-series inputs, RNNs have a memory cell and neurons (or gates) in each layer. The most commonly used RNN is Long Short Term Memory (LSTM)~\cite{lstm} that is used for natural language processing. LSTM network is capable of remembering the close-by and far-apart words from a given word in a sentence. After the training phase, the model is able to predict the next set of words given a history of previous words. 

\subsubsection{\textbf{Long Short Time Memory (LSTM)}}
LSTM~\cite{lstm} is the most widely used recurrent neural network. \textit{Input gate}, \textit{Output gate} and \textit{Forget gate} are the three types of gates used in LSTM. The gates in LSTM enable the network to forget or remember a cell state which in turn allow it to model long-term dependencies. We used a LSTM model that predicts the next bitcoin stock price based on the past two days' stock prices. The model receives two stock price values in text and returns the predicted next stock price in text. 

\subsubsection{\textbf{Gated Recurrent Unit (GRU)}}
GRU~\cite{gru} is a variation of LSTM which aims at solving the vanishing gradient problem. \textit{Reset gate} and \textit{Update gate} are the two types of gates in GRU. GRU combines \textit{Forget gate} and \textit{Input gate} into a single \textit{Update gate}. The resulting model is simpler than standard LSTM models. GRU has lesser gates and thus, lesser computational time. We used a pre-trained model that also projects bitcoin stock price like LSTM model.
\vspace{-10pt}
\subsection{Input Data, Pre-trained Models, and Code Structure} \label{sec.input}
Table~\ref{tab.input} shows the sources of pre-trained models, input data, and output data for individual networks. All models have around 90\% or above prediction accuracy.

We implemented each layer as one or two CUDA or OpenCL kernels. To provide correct input to each layer, we partitioned the pre-trained model files into weight files of individual layers. Each CUDA and OpenCL kernel is fed with the corresponding per-layer weight file as input data. The per-layer weight files are enclosed in the benchmark suite repository. We will provide a script file that collects per-layer weight values, which will help researchers also test the neural network with their pre-trained models.

The CUDA kernel configurations of individual networks, such as layer types, gridDim, blockDim, register count, shared memory usage, and constant memory usage, are shown in Table~\ref{tab.network.config}. We assigned one thread per neuron. If a layer uses more neurons than the maximum threads allowed by the target GPU, we ran the layer over multiple kernels. For example, the first convolution layer of AlexNet is executed over four kernels each with 96 thread blocks of 32x32, 32x23, 23x32, and 23x23 threads. ResNet configuration is shown for the first 24 layers out of 50 layers due to the limited space. The OpenCL code of CifarNet and AlexNet used the same configurations with the CUDA kernels, thus not shown in this table.

\subsection{Target Platforms}
CUDA kernels of the proposed benchmark suite are tested on NVIDIA GPU platforms (GK210 and TX1) and a GPU architecture simulator (GPGPU-Sim). OpenCL kernels are tested on an embedded Xilinx FPGA platform (PynQ-Z1). The OpenCL kernels were converted into RTL code by using Vivado High-Level Synthesis (HLS) design suite and deployed on the FPGA board. 
\vspace{-10pt}

\begin{table}[]
\centering
\scalebox{0.75}
{
\begin{tabular}{llll}\hline
              & \textbf{Server}                                                                     & \textbf{Mobile}  & \textbf{Simulator}                                                                              \\ \hline
Architecture  & Kelper GK210                                                         & Maxwell Tegra X1           & Pascal GP102                                                         \\
\# CUDA cores & 2880                                                                       & 256    & 3584                                                                             \\
Global memory & 24 GB                                                                 & 4 GB     & 11 GB                                                                     \\
Shared/L1D     & 128 KB per Block                                                           & 48KB   &   96KB \begin{tabular}[c]{@{}l@{}}(L1D: 64KB (default),\\128KB, 256KB)\end{tabular}                                                             \\
Register      & 65536 per SM                                                            & 32768 & 65536
\\
OS            & Ubuntu 14.04.1                                                             & Ubuntu 14.04.3 LTS     &                                                            \\
CPU           & \begin{tabular}[c]{@{}l@{}}Intel¢ç Xeon¢ç E5¡©2623 \\ 3.0 GHz\end{tabular} & \begin{tabular}[c]{@{}l@{}}ARM¢ç Cortex¢ç-A57\\ 1.9 GHz\end{tabular} & \\ 
Warp scheduler & & & gto (default), lrr, tlv \\\hline

\end{tabular}\vspace{-15pt}
}	\caption{GPU architectures used for evaluation}\label{tab:sim_param}
\end{table}

\begin{table*}[t]
\centering
\scalebox{0.7}
{
\begin{tabular}{cllllll|cllllll}
\hline
\multicolumn{1}{l}{\textbf{Network}}  & \textbf{Layers}   & \textbf{gridDim} & \textbf{blockDim} & \textbf{regs} & \textbf{smem} & \textbf{cmem} & \textbf{Network}                      & \textbf{Layers}   & \textbf{gridDim} & \textbf{blockDim} & \textbf{regs} & \textbf{smem} & \textbf{cmem} \\ \hline
\textbf{GRU}                          & GRU Layer         & (1, 1, 1)        & (10, 10, 1)       & 12            & 504           & 56            & \textbf{LSTM}                         & LSTM Layer        & (1, 1, 1)        & (100, 1, 1)       & 22            & 936           & 60            \\ \hline
\multirow{4}{*}{\textbf{CifarNet}}    & Conv 1            & (1, 1, 1)        & (32, 32, 1)       & 19            & 40            & 16            & \multirow{4}{*}{\textbf{CifarNet (Cont'd)}}    & Conv 3            & (1, 1, 1)        & (32, 32, 1)       & 12            & 40            & 4             \\
                                      & Pool 1            & (1, 1, 1)        & (32, 32, 1)       & 14            & 60            & 20            &                                       & Pool 3            & (1, 1, 1)        & (32, 32, 1)       & 14            & 60            & 20            \\
                                      & Conv 2            & (1, 1, 1)        & (32, 32, 1)       & 21            & 56            & 16            &                                       & FC 1              & (1, 1, 1)        & (64, 1, 1)        & 19            & 40            & 16            \\
                                      & Pool 2            & (1, 1, 1)        & (32, 32, 1)       & 8             & 40            & 4             &                                       & FC 2              & (1, 1, 1)        & (32, 1, 1)        & 10            & 60            & 12            \\ \hline
\multirow{9}{*}{\textbf{AlexNet}}     & Conv 1-1 & (96,1,1) & (32,32,1) & 19 & 56 & 208     & \multirow{9}{*}{\textbf{AlexNet (Cont'd)}} & Norm & (256,1,1) & (27,27,1) & 13 & 60 & 308 \\
                                      & Conv 1-2 & (96,1,1) & (32,23,1) & 19 & 56 & 208     & & Pool & (256,1,1) & (13,13,1) & 12 & 60 & 204 \\
                                      & Conv 1-3 & (96,1,1) & (23,32,1) & 19 & 56 & 208     & & Conv & (384,1,1) & (13,13,1) & 18 & 80 & 204 \\
                                      & Conv 1-4 & (96,1,1) & (23,23,1) & 19 & 56 & 208     & & Conv 4-1 & (192,1,1) & (13,13,1) & 18 & 80 & 204 \\
                                      & Norm 1-1 & (96,1,1) & (32,32,1) & 13 & 64 & 308    & & Conv 4-2 & (192,1,1) & (13,13,1) & 19 & 80 & 204 \\
                                      & Norm 1-2 & (96,1,1) & (32,23,1) & 13 & 64 & 308    & & Conv 5-1 & (128,1,1) & (13,13,1) & 18 & 80 & 204 \\
                                      & Norm 1-3 & (96,1,1) & (23,32,1) & 13 & 64 & 308    & & Conv 5-2 & (128,1,1) & (13,13,1) & 19 & 80 & 204 \\
                                      & Norm 1-4 & (96,1,1) & (23,23,1) & 13 & 64 & 308    & & Pool & (256,1,1) & (6,6,1) & 12 & 60 & 204 \\
                                      & Pool & (96,1,1) & (27,27,1) & 12 & 60 & 204         & & FC 1 & (4096,1,1) & (1,1,1) & 8 & 58 & 204 \\ 
                                      & Conv 2-1 & (128,1,1) & (27,27,1) & 18 & 80 & 204        & & FC 2 & (4096,1,1) & (1,1,1) & 8 & 58 & 204 \\
                                      & Conv 2-2 & (128,1,1) & (27,27,1) & 18 & 80 & 204        & & FC 3 & (1000,1,1) & (1,1,1) & 8 & 58 & 204 \\ \hline

\multirow{15}{*}{\textbf{SqueezeNet}} & Conv 1            & (111,1,1)        & (111,1,1)         & 19            & 56            & 12            & \multirow{15}{*}{\textbf{SqueezeNet(Cont'd)}} & Fire6\_Squeeze1x1 & (27,1,1)         & (27,1,1)          & 13            & 40            & 0             \\
                                      & Max Pool          & (111,1,1)        & (111,1,1)         & 21            & 40            & 20            &                                       & Fire6\_Expand1x1  & (27,1,1)         & (27,1,1)          & 21            & 40            & 20            \\
                                      & Fire2\_Sqeeze1x1  & (55,1,1)         & (55,1,1)          & 15            & 40            & 4             &                                       & Fire6\_Expand3x3  & (27,1,1)         & (27,1,1)          & 21            & 40            & 20            \\
                                      & Fire2\_Expand1x1  & (55,1,1)         & (55,1,1)          & 13            & 40            & 0             &                                       & Fire7\_Squeeze1x1 & (27,1,1)         & (27,1,1)          & 12            & 60            & 12            \\
                                      & Fire2\_Expand1x1  & (55,1,1)         & (55,1,1)          & 21            & 40            & 20            &                                       & Fire7\_Expand1x1  & (27,1,1)         & (27,1,1)          & 21            & 40            & 20            \\
                                      & Fire3\_Squeeze1x1 & (55,1,1)         & (55,1,1)          & 13            & 40            & 0             &                                       & Fire7\_Expand3x3  & (27,1,1)         & (27,1,1)          & 15            & 40            & 4             \\
                                      & Fire3\_Expand1x1  & (55,1,1)         & (55,1,1)          & 13            & 40            & 0             &                                       & Fire8\_Squeeze1x1 & (27,1,1)         & (27,1,1)          & 13            & 40            & 0             \\
                                      & Fire3\_Expand3x3  & (55,1,1)         & (55,1,1)          & 13            & 40            & 0             &                                       & Fire8\_Expand1x1  & (27,1,1)         & (27,1,1)          & 13            & 40            & 0             \\
                                      & Fire4\_Squeeze1x1 & (55,1,1)         & (55,1,1)          & 13            & 40            & 0             &                                       & Fire8\_Expand3x3  & (27,1,1)         & (27,1,1)          & 13            & 40            & 0             \\
                                      & Fire4\_Expand1x1  & (55,1,1)         & (55,1,1)          & 12            & 60            & 12            &                                       & Max Pool          & (27,1,1)         & (27,1,1)          & 12            & 60            & 12            \\
                                      & Fire4\_Expand3x3  & (55,1,1)         & (55,1,1)          & 13            & 40            & 0             &                                       & Fire9\_Squeeze1x1 & (13,1,1)         & (13,1,1)          & 21            & 40            & 20            \\
                                      & Max Pool          & (55,1,1)         & (55,1,1)          & 13            & 40            & 0             &                                       & Fire9\_Expand1x1  & (13,1,1)         & (13,1,1)          & 13            & 40            & 0             \\
                                      & Fire5\_Squeeze1x1 & (27,1,1)         & (27,1,1)          & 13            & 40            & 12            &                                       & Fire9\_Expand3x3  & (13,1,1)         & (13,1,1)          & 9             & 32            & 12            \\
                                      & Fire5\_Expand1x1  & (27,1,1)         & (27,1,1)          & 21            & 40            & 20            &                                       & Conv 10           & (15,1,1)         & (15,1,1)          & 13            & 40            & 0             \\
                                      & Fire5\_Expand3x3  & (27,1,1)         & (27,1,1)          & 21            & 40            & 20            &                                       & Global Avg Pool   & (1,1,1)          & (1000,1,1)        & 14            & 40            & 0             \\ \hline
\multirow{12}{*}{\textbf{ResNet}}     & Conv              & (64,1,1)         & (32,32,1)         & 29            & 76            & 12            & \multirow{12}{*}{\textbf{ResNet (Cont'd)}}     & Conv              & (64,1,1)         & (32,32,1)         & 31            & 84            & 8             \\
                                      & BatchNorm         & (64,1,1)         & (32,32,1)         & 12            & 52            & 12            &                                       & BatchNorm         & (64,1,1)         & (32,32,1)         & 12            & 52            & 12            \\
                                      & Scale             & (64,1,1)         & (32,32,1)         & 12            & 52            & 4             &                                       & Scale             & (64,1,1)         & (32,32,1)         & 12            & 52            & 4             \\
                                      & Relu              & (64,1,1)         & (32,32,1)         & 8             & 32            & 8             &                                       & Relu              & (64,1,1)         & (32,32,1)         & 8             & 32            & 8             \\
                                      & Pool              & (64,1,1)         & (32,32,1)         & 19            & 68            & 4             &                                       & Conv              & (256,1,1)        & (32,32,1)         & 31            & 84            & 8             \\
                                      & Conv              & (256,1,1)        & (32,32,1)         & 31            & 84            & 8             &                                       & BatchNorm         & (256,1,1)        & (32,32,1)         & 12            & 52            & 12            \\
                                      & BatchNorm         & (256,1,1)        & (32,32,1)         & 5             & 48            & 12            &                                       & Scale             & (256,1,1)        & (32,32,1)         & 12            & 52            & 4             \\
                                      & Scale             & (256,1,1)        & (32,32,1)         & 12            & 52            & 4             &                                       & Eltwise           & (256,1,1)        & (32,32,1)         & 11            & 48            & 4             \\
                                      & Conv              & (64,1,1)         & (32,32,1)         & 31            & 84            & 8             &                                       & Relu              & (256,1,1)        & (32,32,1)         & 8             & 32            & 8             \\
                                      & BatchNorm         & (64,1,1)         & (32,32,1)         & 12            & 52            & 12            &                                       & Conv              & (64,1,1)         & (32,32,1)         & 31            & 84            & 8             \\
                                      & Scale             & (64,1,1)         & (32,32,1)         & 12            & 52            & 4             &                                       & BatchNorm         & (64,1,1)         & (32,32,1)         & 12            & 52            & 12            \\
                                      & Relu              & (64,1,1)         & (32,32,1)         & 8             & 32            & 8             &                                       & Scale             & (64,1,1)         & (32,32,1)         & 12            & 52            & 4             \\ \hline
\multirow{9}{*}{\textbf{VGGNet}} & Conv & (16,16,64) & (14,14,1) & 15 & 0 & 72 & \multirow{9}{*}{\textbf{VGGNet (Cont'd)}} & Pool & (7,7,256) & (4,4,1) & 13 & 0 & 56 \\
                        & Conv & (16,16,64) & (14,14,1) & 19 & 0 & 76 &                         & Conv & (7,7,512) & (4,4,1) & 19 & 0 & 76 \\
                        & Pool & (8,8,64)   & (14,14,1) & 13 & 0 & 56 &                         & Conv & (7,7,512) & (4,4,1) & 19 & 0 & 76 \\
                        & Conv & (8,8,128)  & (14,14,1) & 19 & 0 & 76 &                         & Conv & (7,7,512) & (4,4,1) & 19 & 0 & 76 \\
                        & Conv & (8,8,128)  & (14,14,1) & 19 & 0 & 76 &                         & Pool & (7,7,512) & (2,2,1) & 13 & 0 & 56 \\
                        & Pool & (8,8,128)  & (7,7,1)   & 13 & 0 & 56 &                         & Conv & (7,7,512) & (2,2,1) & 19 & 0 & 76 \\
                        & Conv & (8,8,256)  & (7,7,1)   & 19 & 0 & 76 &                         & Conv & (7,7,512) & (2,2,1) & 19 & 0 & 76 \\
                        & Conv & (8,8,256)  & (7,7,1)   & 19 & 0 & 76 &                         & FC   & (4,4,4)   & (8,8,1) & 11 & 0 & 77 \\
                        & Conv & (8,8,256)  & (7,7,1)   & 19 & 0 & 76 &                         & FC   & (1,1,10)  & (10,10,1) & 11  & 0  & 77  \\ \hline
\end{tabular}
}\caption{Network Configuration and SRAM Usage: deep networks are laid across two columns (For ResNet, the first 24 layers are shown due to the space limitation) } \label{tab.network.config}\vspace{-15pt}
\end{table*}

\section{Characterizations and Analysis}\label{sec:Results}
Most of the detailed statistics have been measured by using GPGPU-Sim while varying various architecture configurations such as cache size and warp scheduler. A few runtime system statistics (i.e. on-chip memory usage and power consumption) were measured on real devices. For the power measurement, we used GPUWattch~\cite{gpuwattch} for detailed statistics and Wattsup power meter for device-level statistics. For better understanding, we marked the evaluated platform to each graph if the evaluation was not measured with GPGPU-Sim. The configurations of the evaluated GPU and FPGA architectures are specified in Table~\ref{tab:sim_param} and Table~\ref{tab:sim_param.fpga}, respectively. We used the development branch GPGPU-Sim because it supports Pascal architecture, which is one of the latest architectures.

\begin{table}[]
\centering
\scalebox{0.75}
{
\begin{tabular}{ll}\hline
              & \textbf{Mobile FPGA}            \\ \hline
Processor  & Dual-core ARM Cortex-A9 @ 650 MHz              \\
Memory & 512MB DDR3\\
Storage & 32 GB \\
Programmable Logic & Xilinx Zynq Z7020\\
& 13,300 logic slices\\
& 630KB BRAM \\\hline

\end{tabular}\vspace{-15pt}
}	\caption{FPGA platform used for evaluation}\label{tab:sim_param.fpga}
\end{table}
\vspace{-10pt}
\subsection{Overall Performance}\label{eval.perf}

\subsubsection{Per-Layer Execution Time}

\begin{figure}[htb]
\centering
\includegraphics[width=0.50\textwidth]{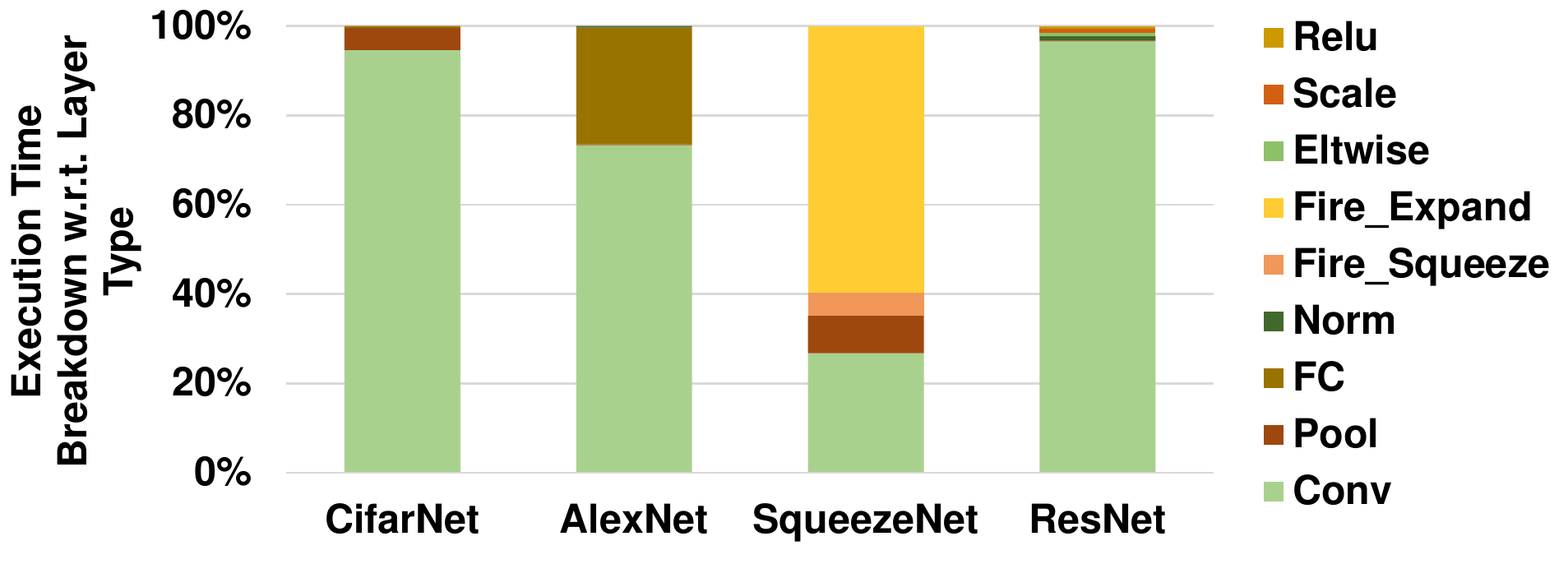}
\caption{Execution Time Breakdown w.r.t. Layer Type} \label{fig.exe.breakdown}
\end{figure}

We first measured the execution time contribution of individual layer types. As each network uses different number and size of layers, it is meaningless to compare the total execution time across networks. Instead, it would be more insightful to know the types of layers that consume the most time for each network. Figure~\ref{fig.exe.breakdown} shows the execution time breakdown with regard to layer types of CNNs. Note that RNNs run the same type layer repeatedly, we evaluated only CNNs for this experiment. As can be seen in the figure, convolution layers take the majority of the execution time of all networks. Especially in CifarNet and ResNet, over 90\% of execution time is spent by convolution layers. In SqueezeNet, though fire modules are yet another convolution layers, we separatedly evaluated them from the regular convolution layers according to the network specification~\cite{squeezenet}. It is observed that the fire\_expand layers take more time than convolution layers in SqueezeNet because eight times more fire\_expand layers are executed than convolution layers. However, according to our evaluation, the longest layer of SqueezeNet is still the last convolution layer (conv 10), which is 40\% longer than the longest fire\_expand layer. 

\textit{\underline{Observation 1.} Convolution layer is the most time-consuming layer of CNNs, which may be the best target for optimization.}

\subsubsection{Performance Impact of On-chip Cache}

\begin{figure}[htb]
\centering
\includegraphics[width=0.50\textwidth]{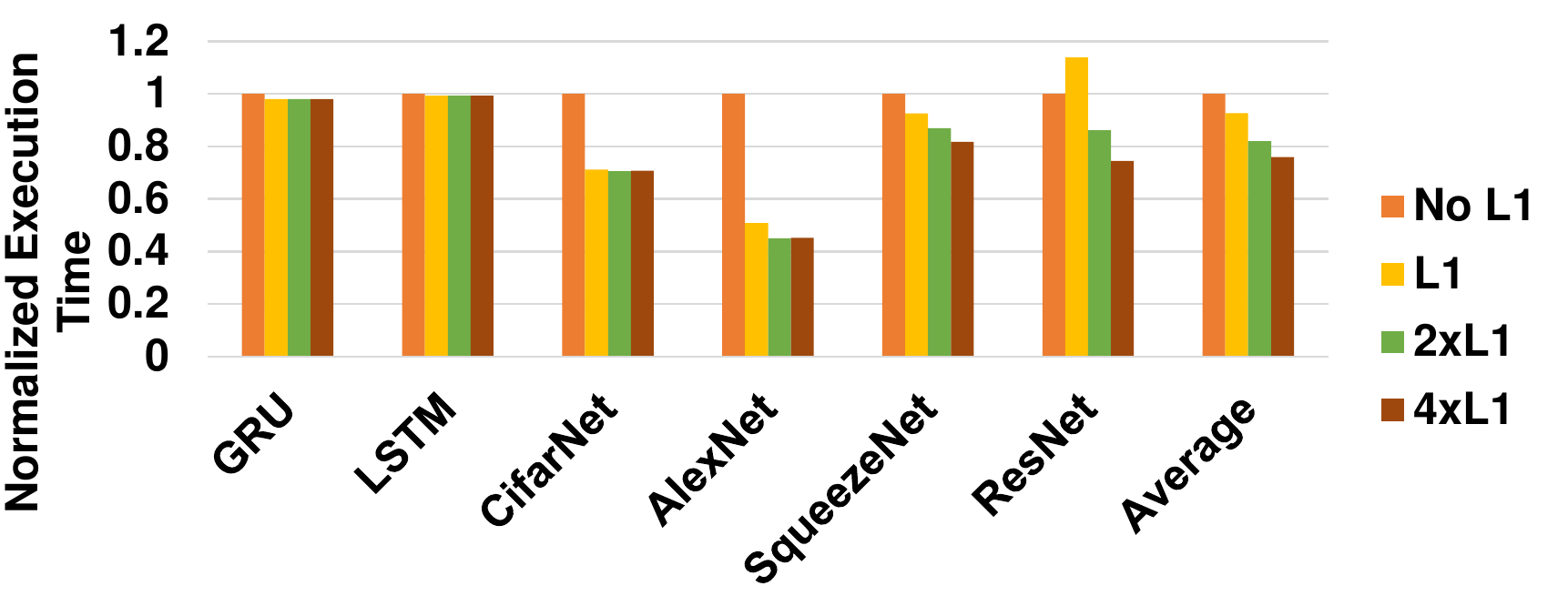}
\caption{Normalized Execution Time with Various L1D Sizes} \label{fig.exe.l1} 
\end{figure}

Our next evaluation related to performance is the on-chip cache sensitivity. The impact of L1D cache has not been well studied for RNNs. For CNNs, it is known that L1D is too small to provide better performance for large CNNs because of cache thrashing overhead. However, there has not been a study that shows the performance impact of various L1D size because most of the existing studies have been evaluated on real GPUs, which is hard to reconfigure the cache size. We evaluated RNNs and CNNs on GPGPU-Sim while varying L1D size from zero to 4$\times$64KB as shown in Figure~\ref{fig.exe.l1}. Note that 64KB is the default L1D size of Pascal architecture. The total execution times with three different size L1Ds are normalized by the execution time of when L1D is bypassed (marked as $No\ L1$). RNNs do not show performance improvement with larger L1Ds. It is an expected result because RNNs use relatively small input data (i.e., a few stock price values or a series of words) and there is not much repeatedly accessed data. On the other hand, most of the CNNs show a significant performance improvement with larger on-chip caches. For example, AlexNet shows 2$\times$ speedup with a 64KB L1D and the performance is further improved by 10\% with 2$\times$ larger L1D. It is also an expected result because CNNs inherently have a lot of redundant data accesses. For example, the same convolution feature maps are used by all neurons in the same layer and neighboring neurons use overlapping input data. With the great performance improvement by CNNs, the execution time is reduced by 10\% on average when employing 64KB larger L1Ds across the networks.  

\textit{\underline{Observation 2.} On-chip cache is helpful for improving the performance of CNNs while the impact of on-chip cache for RNNs is negligible.}
\vspace{-10pt}
\subsection{Power Consumption}

\begin{figure}
\center
\includegraphics[width=0.50\textwidth]{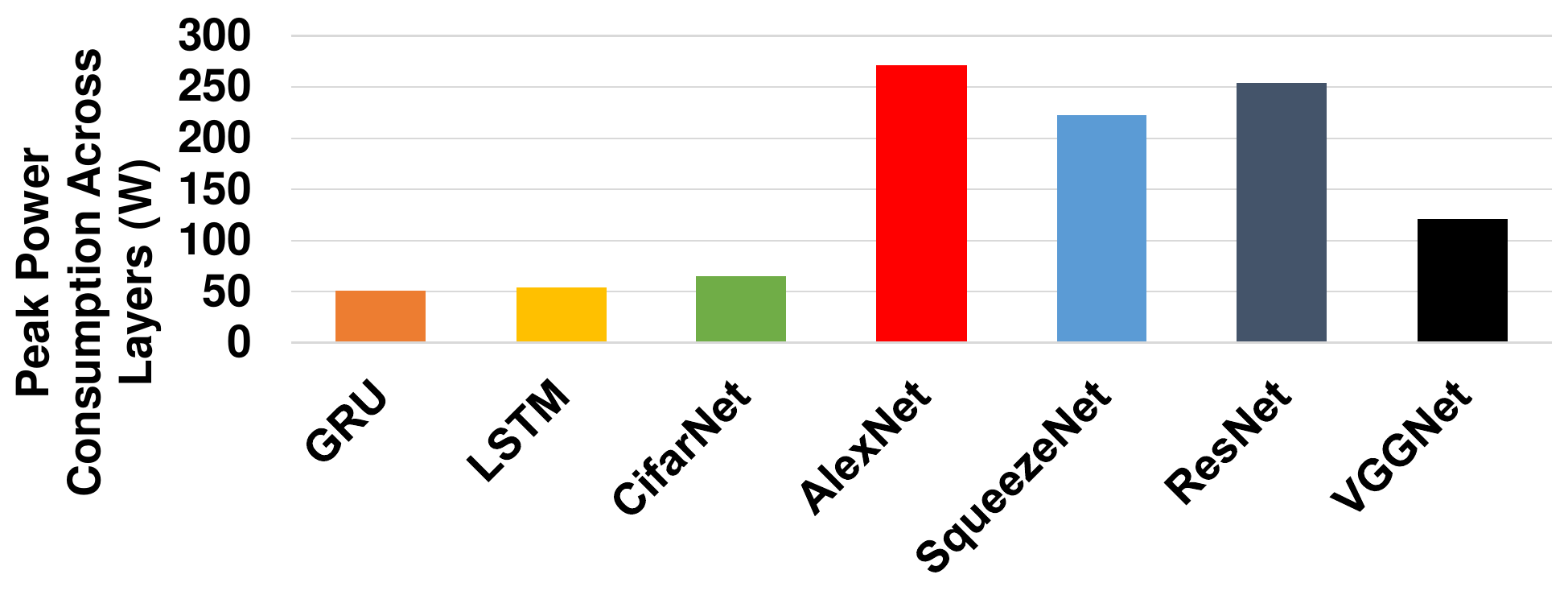}
\caption{Peak Power Consumption Across Layers in Watt} \label{fig.peak.power}\vspace{-15pt}
\end{figure}
Again, as networks use different number of layers, it is meaningless to compare the total energy consumption. Thus, we checked peak power consumption, the most power hungry layer type, and the most power hungry micro-architecture components while running the networks. 

\subsubsection{Peak Power Consumption}

Figure~\ref{fig.peak.power} shows the maximum power consumption that was ever measured during individual network executions. The peak power consumption showed correlation with the size of layers. For example, AlexNet and ResNet that use larger layers derived the higher peak power consumption. As shown in Table~\ref{tab.network.config}, AlexNet runs 128 thread groups of 27$\times$27 threads for some convolution layers, which is over 100 times larger than CifarNet's largest convolution layer size. Such a layer size disparity leads to 5 times higher peak power consumption of AlexNet than CifarNet. 

\textit{\underline{Observation 3.} Neural networks that use larger layers (more neurons) typically show higher peak power.}

\subsubsection{Per-layer Power Consumption}

\begin{figure}
\center
\includegraphics[width=0.50\textwidth]{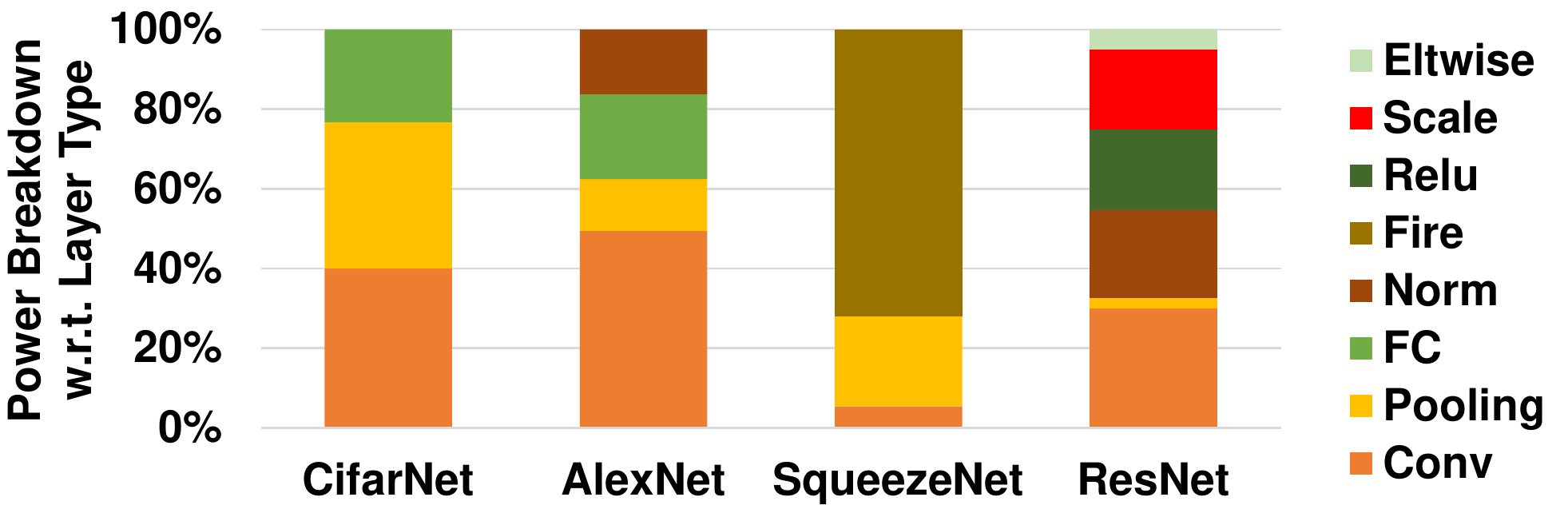}
\caption{Average Power Consumption per Layer Type in Watt} \label{fig.power.breakdown.layer}
\end{figure}

\begin{figure}
\center
\includegraphics[width=0.50\textwidth]{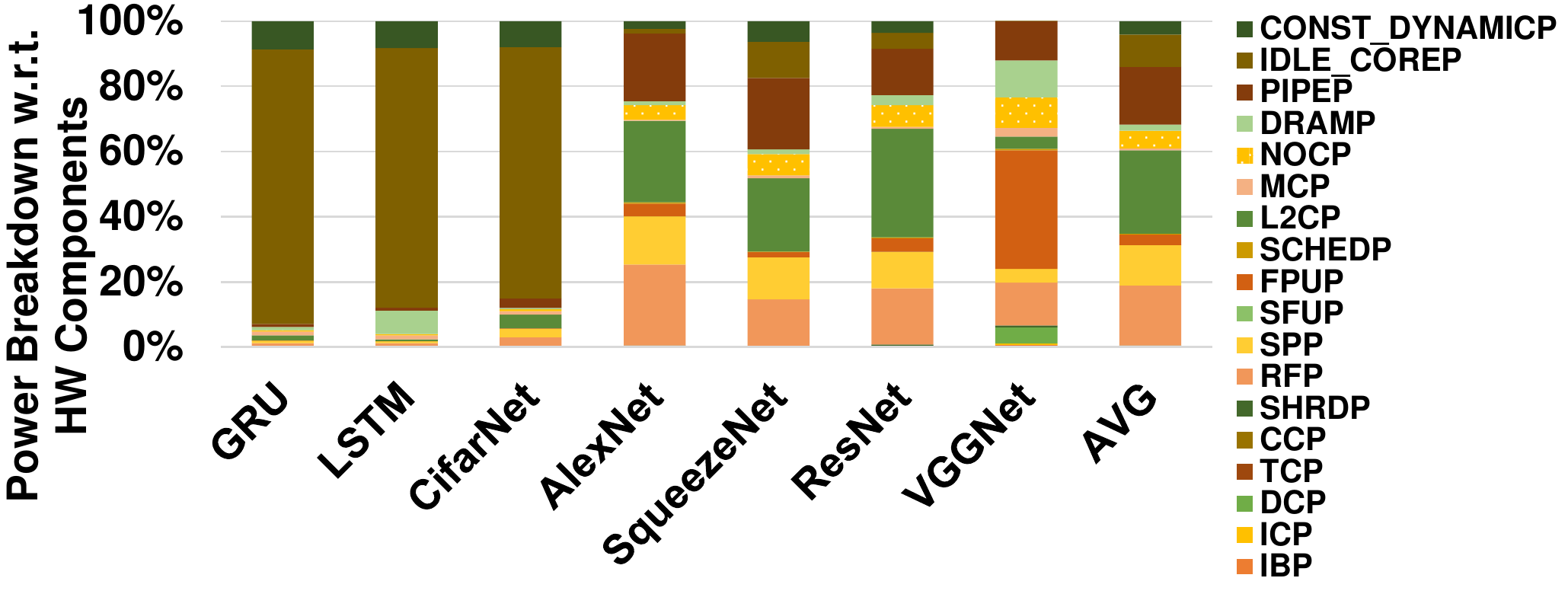}
\caption{Breakdown of Average Power Consumption} \label{fig.power.breakdown}\vspace{-5pt}
\end{figure}

To understand the power hungry layers among various layer types, we broke down the total power consumption with regard to the kinds of layers. Though convolution layers are the incomparably time-consuming layers as shown in Figure~\ref{fig.exe.breakdown}, power consumption is relatively more balanced. As shown in Figure~\ref{fig.power.breakdown.layer}, pooling layers consume almost similar amount of power with convolution layers in CifarNet. Also, the combined power dissipation of Scale, Relu, and Norm layers is higher than that of convolution layers in ResNet. 

To understand this balanced power dissipation, we also measured power breakdown with regard to micro-architecture components as shown in Figure~\ref{fig.power.breakdown}. The key power consumers are register file (marked as $RFP$), L2 cache (marked as $L2CP$), and idle core power (marked as $IDLE\_COREP$). We observed that the register usage across layers is not significantly different as shown in Table~\ref{tab.network.config}. Thus, we believe that the main driver of the balanced power distribution is the L2 cache accesses. The total L2 misses of different layers show a correlated statistics as shown in Figure~\ref{fig.l2.miss.layer}. Figure~\ref{fig.l2.miss.layer} plots the total number of L2 misses per layer type when L1D is bypassed. In CifarNet, convolution layers encounter similar number of L2 misses with fully connected layers (marked as $FC$). Likely, in AlexNet, fully connected layers show even greater number of L2 misses than convolution layers and the total number of misses of non-convolution layers of ResNet is comparable with that of convolution layers. Consequently, it is believed that non-convolution layers consume as much power as convolution layers because of cache accesses. 

\textit{\underline{Observation 4.} Though convolution layers are the most time-consuming layers, all layers consume similar amount of power due to cache and memory accesses.}

\subsubsection{Power Efficiency of Different Platforms}

\begin{figure}[htb]
\centering
\includegraphics[width=0.38\textwidth]{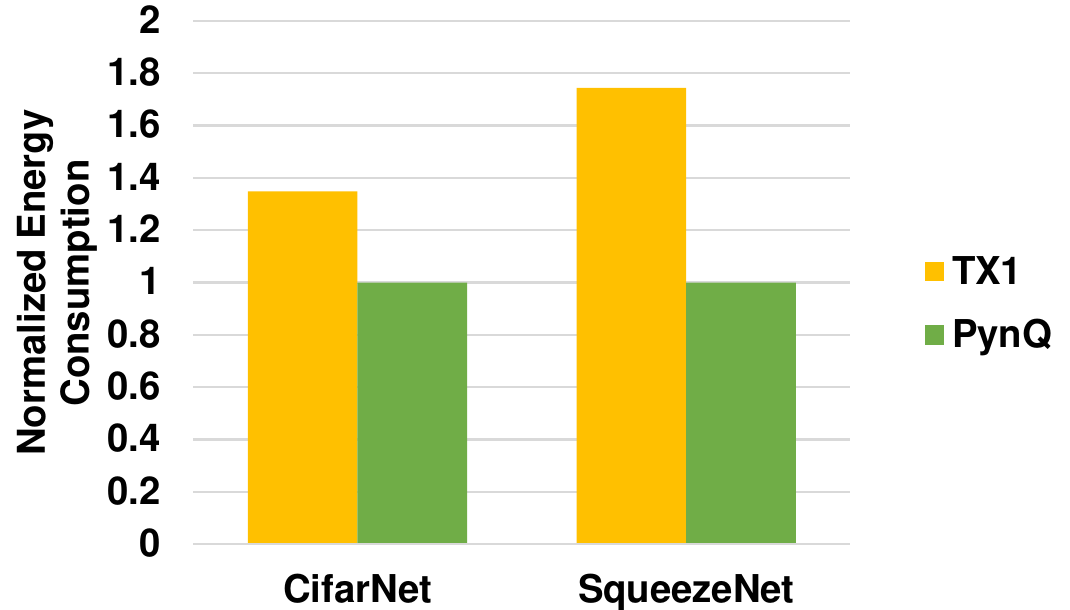}
\caption{Energy Consumption on Embedded GPU (TX1) vs. Embedded FPGA (PynQ) boards} \label{fig.power.platforms}
\end{figure}

We also evaluated energy efficiency of different platforms. Figure~\ref{fig.power.platforms} shows the normalized energy consumption measured on an embedded GPU board (NVIDIA TX1) and an embedded FPGA board (Xilinx PynQ) for CifarNet and SqueezeNet. We measured the peak power consumption by using a Wattsup power meter. As the power meter does not show the total energy consumption, we calculated the energy consumption by multiplying the peak power consumption with the total execution time. TX1 showed 2.28$\times$ and 3.2$\times$ higher peak power consumption than PynQ for CifarNet and SqueezeNet, respectively. This is an expected result because TX1 is equipped with more hardware resources (i.e. larger memory size) and runs general-purpose pipeline while PynQ's pipeline is dedicatedly programmed for each network. However, the execution times of the two networks on TX1 were 1.7$\times$ and 1.8$\times$ shorter than on PynQ because of slower code loading time and smaller on-chip memory size of PynQ. Therefore, the overall energy consumption of the two networks on TX1 was 1.34$\times$ and 1.74$\times$ higher than PynQ. 
\vspace{-10pt}
\subsection{Stall Cycle Breakdown}

\begin{figure*}
\center
\includegraphics[width=0.90\textwidth]{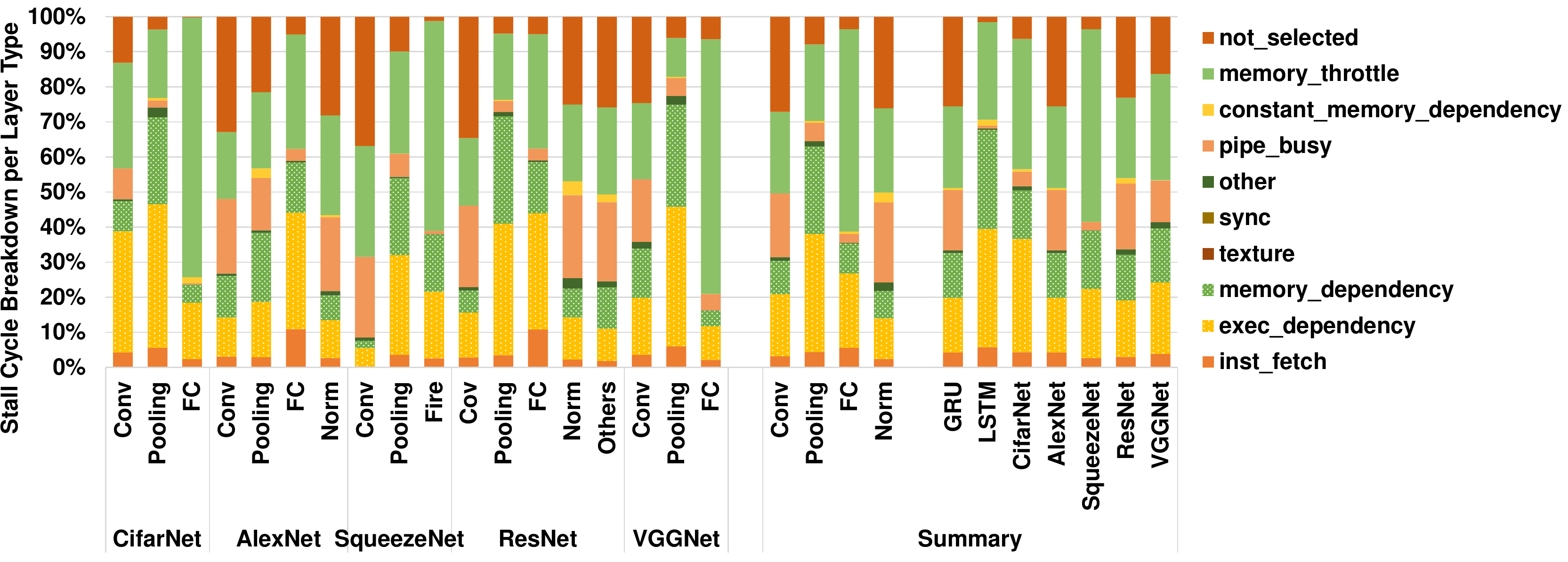}
\caption{Breakdown of Stall Cycles (GK210)} \label{fig.stall.breakdown}
\end{figure*}
To understand performance bottleneck, we collected stall cycles by running a profiler, nvprof~\cite{nvprof}, on an NVIDIA GK210 GPU. Figure~\ref{fig.stall.breakdown} shows stall cycle breakdown of individual layers of each network as well as across networks. As the figure shows, the percentage of stall reasons of each layer type of each network varies. However, there are clear patterns that indicate individual layer types as can be seen from the bar charts in the right-hand side summary section. For example, fully-connected layers suffer from memory throttling more than the other layers. Convolution and normalization layers encounter more stalls due to unavailable pipelines. Pooling layers show higher stall rates due to data dependency than the other layers. These patterns well describe individual layer types. For example, fully-connected layers typically use large data to compute the activation of all features. Thus, the fully-connected layers use higher memory resources such as MSHRs and hence the execution is suspended if all provided memory resources are used up. Convolution and normalization layers typically use more neurons than the other layers, which make the arithmetic operation pipelines busy, thereby throttled by unavailable pipelines. Pooling layers summarize the convolution results either with maximum or average values, which requires repeated comparison of many input data, which leads to high data dependency. 

GRU and LSTM show similar patterns with convolution layers and pooling layers of CNNs, respectively. We believe LSTM encounters more data dependency due to more complex structure than GRU. Note that LSTM uses three gates (input, output, and forget) while GRU uses two (reset and update gates). 

\textit{\underline{Observation 5.} Stall cycle breakdown is a good indicator of layer types, which will be helpful for architecture optimization for each layer type.}
\vspace{-5pt}
\subsection{Instruction Type Characterization}
\subsubsection{The Most Executed Operations}
\begin{figure}
\center
\includegraphics[width=0.50\textwidth]{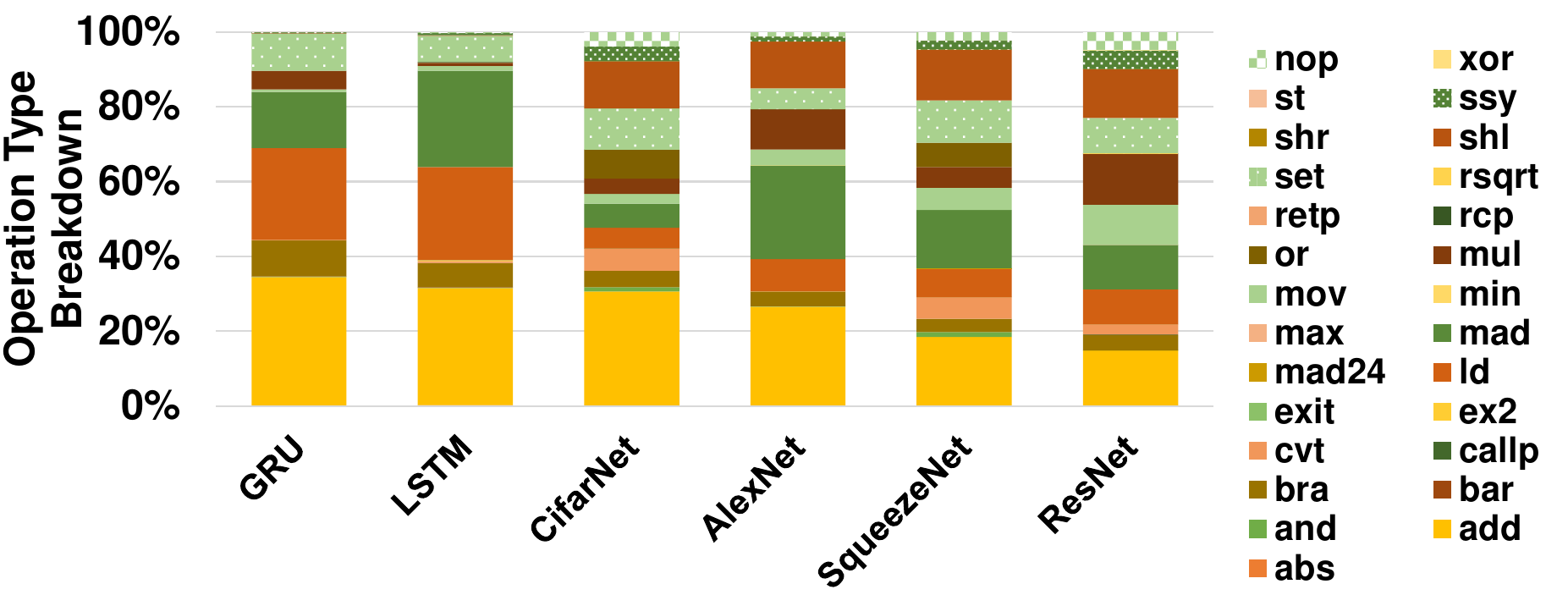}
\caption{Operation Type Breakdown} \label{op_breakdown}\vspace{-15pt}
\end{figure}

\begin{figure}
\center
\includegraphics[width=0.20\textwidth]{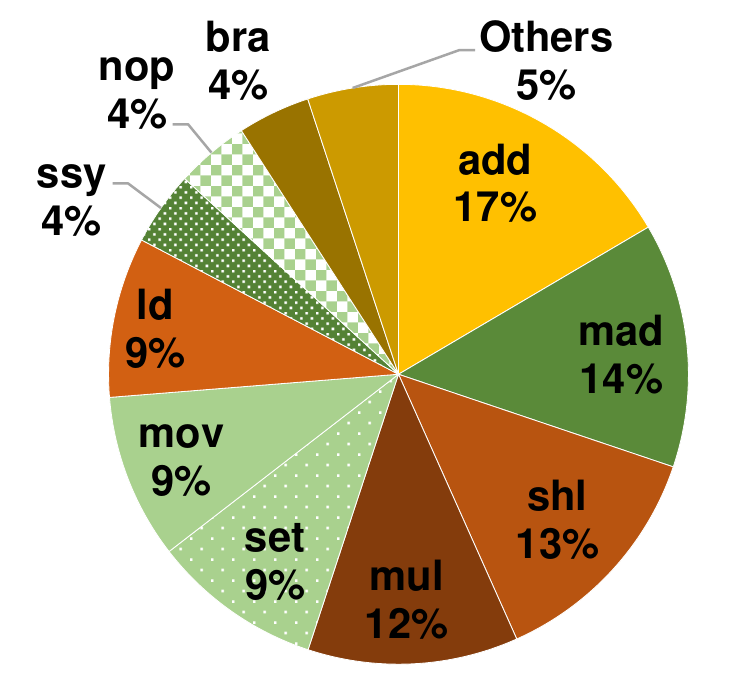}
\caption{Total Operations Breakdown Used By All Networks: Top 10 operations are used for 95\% of execution} \label{op_top_10}\vspace{-15pt}
\end{figure}
 
It is important to understand what kind of instructions are most frequently executed by neural networks. Figure~\ref{op_breakdown} and Figure~\ref{op_top_10} show the operation type breakdown of individual networks and the top 10 operations that are commonly executed by all networks. As can be seen in Figure~\ref{op_breakdown}, GRU and LSTM show similar breakdown while the other four CNNs show another similar breakdown pattern. This is because the internal algorithm of CNN (or RNN) is similar though individual networks use different number and size of layers (or gates) in different orders. RNNs use $add$, $ld$, $mad$, and $set$ instructions the most. Adding to these four instructions, CNNs also use $shl$ and $mul$ excessively. The intensive usage of $add$, $ld$, $mad$ or $mul$ is intuitive because the main algorithm of neural networks is $\sum_{i} w_{i} \times x_{i} + b$ where $w$ is weight, $x$ is input, and $b$ is bias. We found that $shl$ is mainly used for calculating the data index in warp unit (i.e. each warp runs 32 threads thus shift-left-by-5 is used to calculate the warp-unit data accesses). 

Figure~\ref{op_top_10} shows the most executed instructions across all networks. As expected, $add$, $mad$, $mul$, and $shl$ are the most executed instructions where these four instructions are used over 50\% of entire execution. The top 10 instructions shown in the graph are used for 95\% of entire execution. This means that architects can focus on optimizing the pipelines to better support these top instructions. 

\textit{\underline{Observation 6.} Operation breakdown is a good indicator of CNN and RNN.}

\textit{\underline{Observation 7.} Top four mostly executed operations ($add$, $mad$, $mul$, and $shl$) are used for over a half of the entire execution and top 10 operations are used for 95\% of the entire execution.}

\subsubsection{The Most Used Data Types}
\begin{figure}
\center
\includegraphics[width=0.48\textwidth]{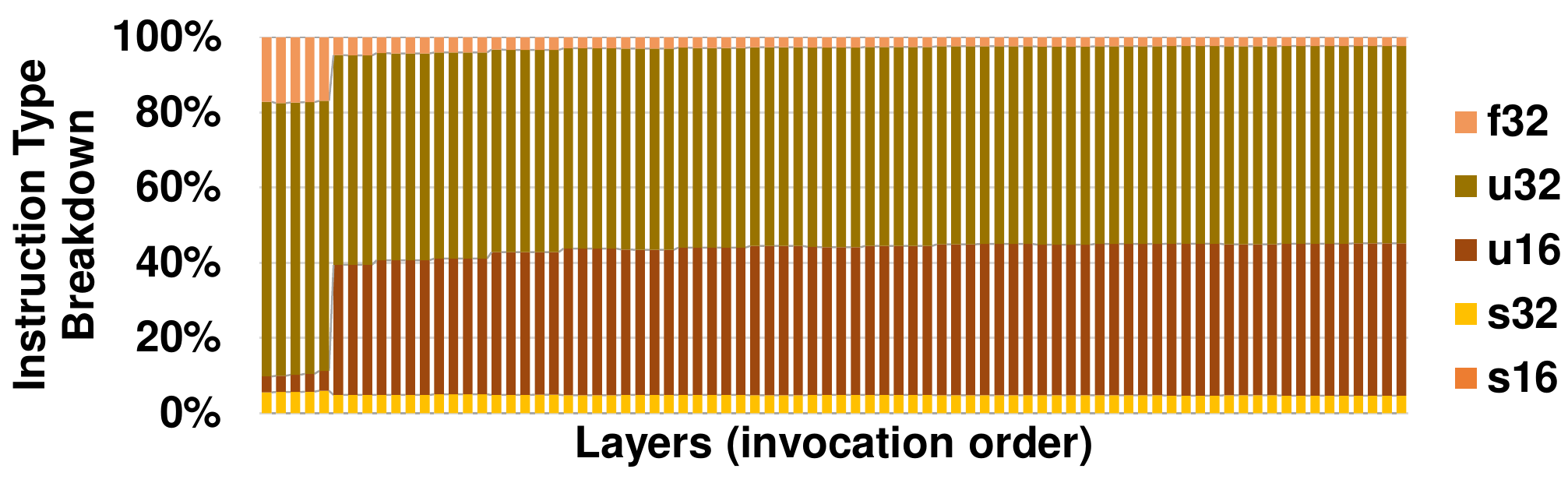}
\caption{Instruction Type Breakdown Throughout Execution: the case of ResNet which is similar to all networks} \label{inst_breakdown}
\end{figure}

To improve performance and power efficiency, quantized networks have been recently introduced~\cite{TPU, bnn}. We plan to apply quantization for the proposed benchmark suite but the current version uses 32-bit floating-point data as inputs. Though input data are 32-bit floating-point values, we would like to understand what other types of data are mostly used by the neural networks. Figure~\ref{inst_breakdown} shows the instruction data type breakdown throughout the execution of ResNet. We observed that the other networks show similar patterns, thus omitted the statistics of the other networks. 

Interestingly, the percentage of 32-bit floating-point (marked as $f32$) was not the dominant data type. In the first a few layers, around 20\% of the total instructions used 32-bit floating-point data. However, in the deeper layers, the portion of 32-bit floating-point is even reduced. Overall, the most used data types are unsigned 32-bit and 16-bit integers. Signed 32-bit integer is the next most frequently used data type. We believe this is mainly because of the activation function, $ReLU$. ReLU resets any negative floating-point value to zeros. Thus, as execution continues, significant amount of data become zero's, which can be handled by integer pipeline. Another reason of high portion of integer values is the intensive data index calculation. While the data that are handled by neural network algorithms are floating-point values, to have all neurons access their data in large input matrixes, considerable amount of index calculation is conducted. Especially in GPUs that run a batch of threads (32 threads) in a lock-step manner (in a warp), the index calculation needs additional operations for warp-unit data accesses.

\textit{\underline{Observation 8.} Even without quantization, neural networks run significant amount of integer operations mainly due to activation function outputs and data index calculation.}

\subsection{Memory Footprint}

\subsubsection{Device Memory Usage}

\begin{figure}[htb]
\centering
\includegraphics[width=0.50\textwidth]{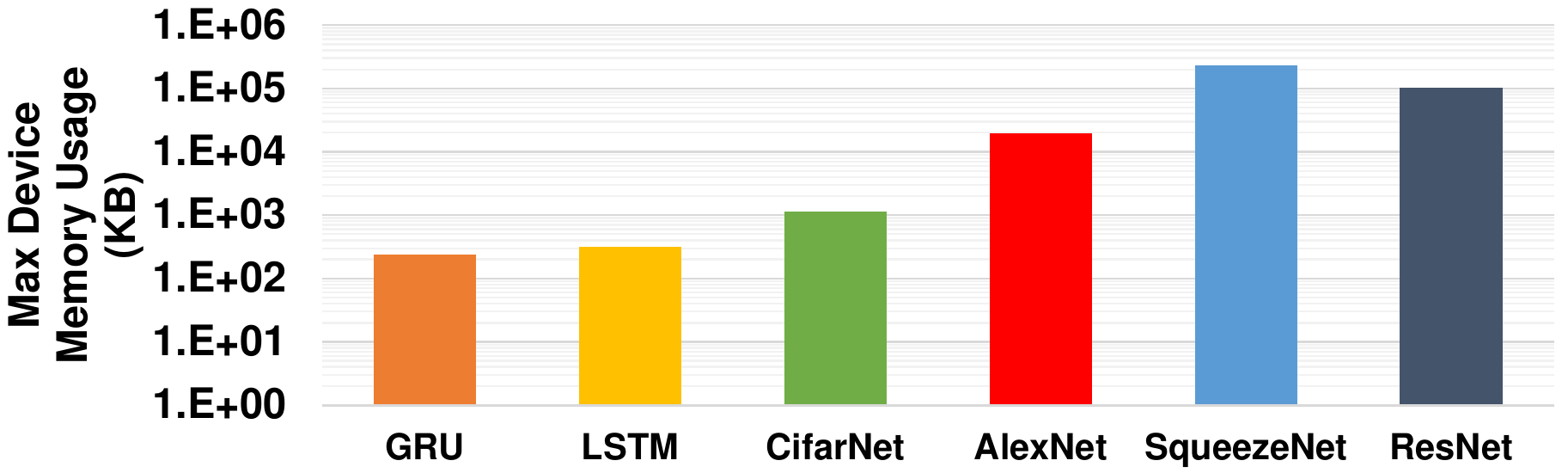}
\caption{Memory Footprint (TX1)} \label{mem.footprint}
\end{figure}

To understand the memory requirement, we measured the device memory usage by using nvprof on TX1. Figure~\ref{mem.footprint} shows the maximum device memory usage while executing all layers of individual networks. The device memory utilization is highly correlated with the pre-trained model size. Thus, the statistics may be different when using different models. However, as many of our benchmark suite code used publicly accessible reference models, the provided size may be considered as a typical model size. GRU and LSTM use less than 500KB memory. Thus, they fit on a small embedded devices such as Xilinx PynQ. However, most of the CNNs use at least 1 MB memory. Thus, in our evaluations on FPGA board had to partition each layer into several sub kernels and run them over multiple iterations.

\textit{\underline{Observation 9.} CNNs have high memory footprint, which needs compression to be deployed on small embedded devices.}

\subsubsection{On-chip Memory Usage}

\begin{figure}[htb]
\centering
\includegraphics[width=0.50\textwidth]{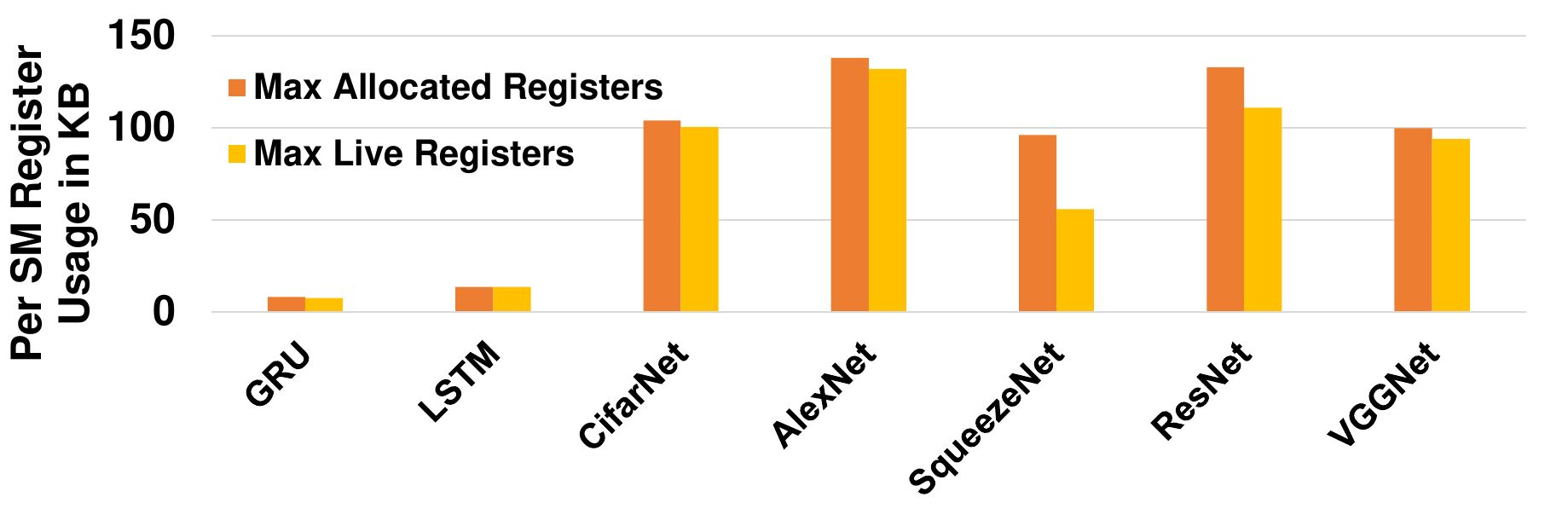}
\caption{Register File Usage in KB} \label{reg.util}
\end{figure}

For GPUs that run massively parallel threads, register files are typically larger than any other on-chip memories. Thus, it is known that the register file is the third most power hungry logic in GPU~\cite{gpuwattch}. Thus, we evaluated the register file utilization of DNNs. Figure~\ref{reg.util} shows the register file utilization measured by running a modified GPGPU-Sim~\cite{micro_hyeran15}, but with a Pascal architecture configuration file. $Max\ Allocated\ Registers$ is the maximum number of registers that are allocated by the compiler and $Max\ Live\ Registers$ is the maximum number of live registers throughout the execution. AlexNet and ResNet use over 50\% of the 256 KB per-SM register file of Pascal architecture, while the live register count is slightly lower than 50\%. However, all the other networks use less than 100 KB registers. Especially RNNs use less than 20KB registers.

\textit{\underline{Observation 10.} Though neural networks are compute-intensive workloads, GPU register file is significantly underutilized.}

\begin{figure}[htb]
\centering
\includegraphics[width=0.50\textwidth]{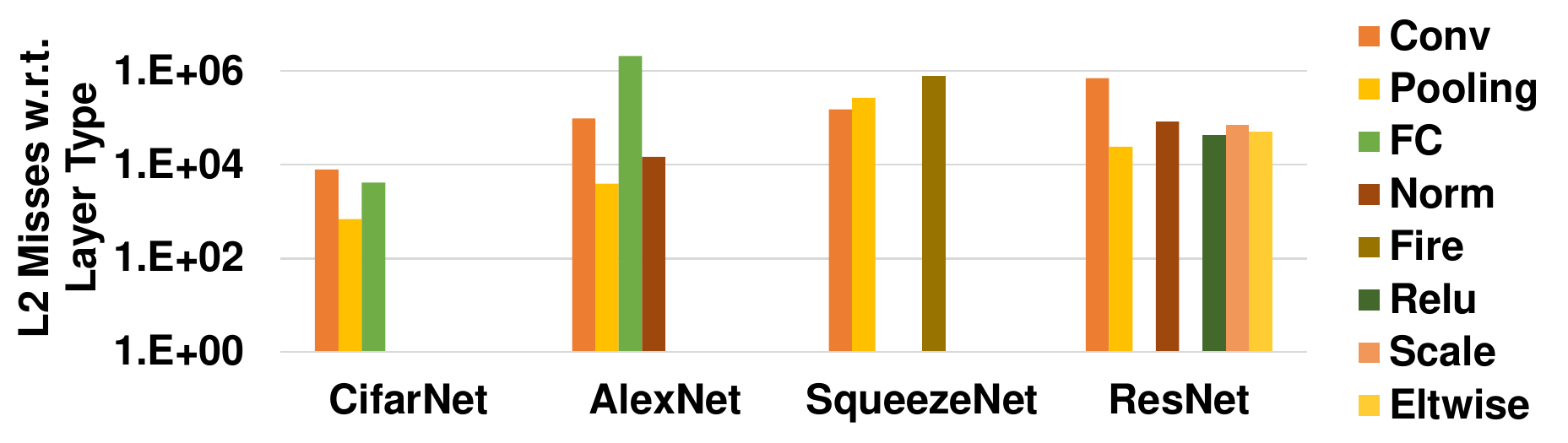}
\caption{Total L2 Misses per Layer Type without L1D} \label{fig.l2.miss.layer}\vspace{-10pt}
\end{figure}

\begin{figure}[htb]
\centering
\includegraphics[width=0.50\textwidth]{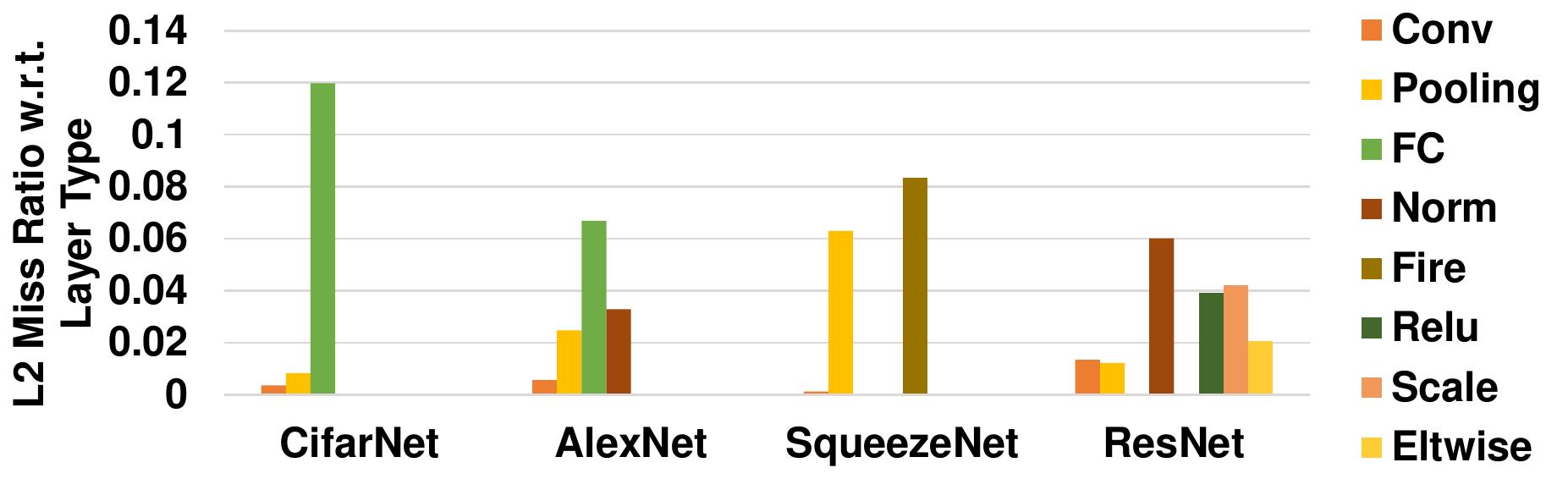}
\caption{L2 Miss Ratio per Layer Type without L1D} \label{fig.l2.miss.ratio.layer}\vspace{-10pt}
\end{figure}

We showed that on-chip cache helps improve performance of CNNs in Section~\ref{eval.perf}. In this section, we evaluate data locality per layer by checking the total number of L2 misses and L2 miss ratio when L1D is not used. Figure~\ref{fig.l2.miss.layer} shows the total number of L2 misses per layer. Clearly, convolution layers and fully-connected layers are the most data intensive layers. However, according to the L2 miss ratio plotted in Figure~\ref{fig.l2.miss.ratio.layer}, convolution layers have significantly lower L2 miss ratio (average of less than 1\%) than fully-connected layers (average of 10\%). This means that convolution layers have higher data locality than fully-connected layers. The high data locality of convolution layer can be found from SqueezeNet and ResNet as well, where convolution layers are one of the top layers that have the most L2 misses but have the lowest L2 miss ratio among all layers. From these statistics, we can conclude that on-chip memory is mainly useful for convolution layers. For the other memory-intensive layers such as fully-connected layers, another optimization technique should be used to reduce the memory overhead. 

\textit{\underline{Observation 11.} Convolution layers have high data locality and hence on-chip memory is mainly useful for optimizing the performance of convolution layers.}

\subsection{Scheduler Sensitivity}

\begin{figure}[htb]
\centering
\includegraphics[width=0.50\textwidth]{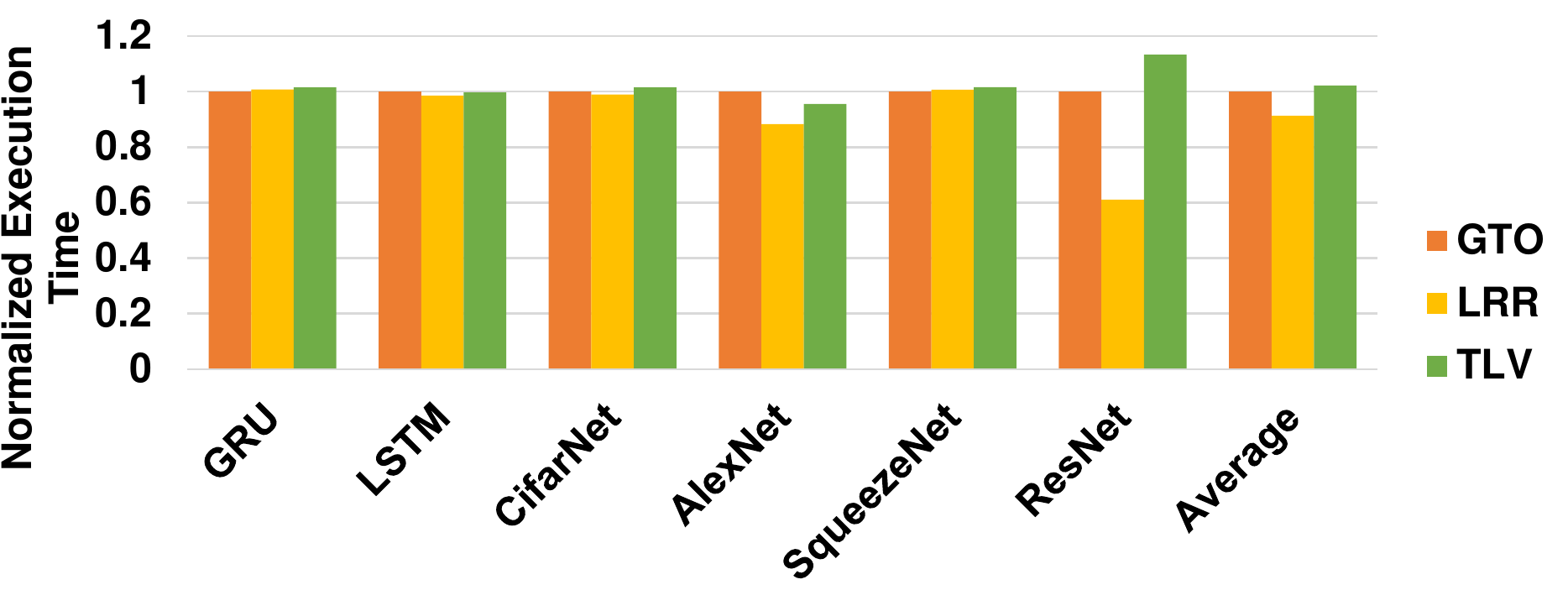}
\caption{Warp Scheduler Sensitivity} \label{sched.sens}\vspace{-15pt}
\end{figure}
\begin{figure}[htb]
\centering
\includegraphics[width=0.50\textwidth]{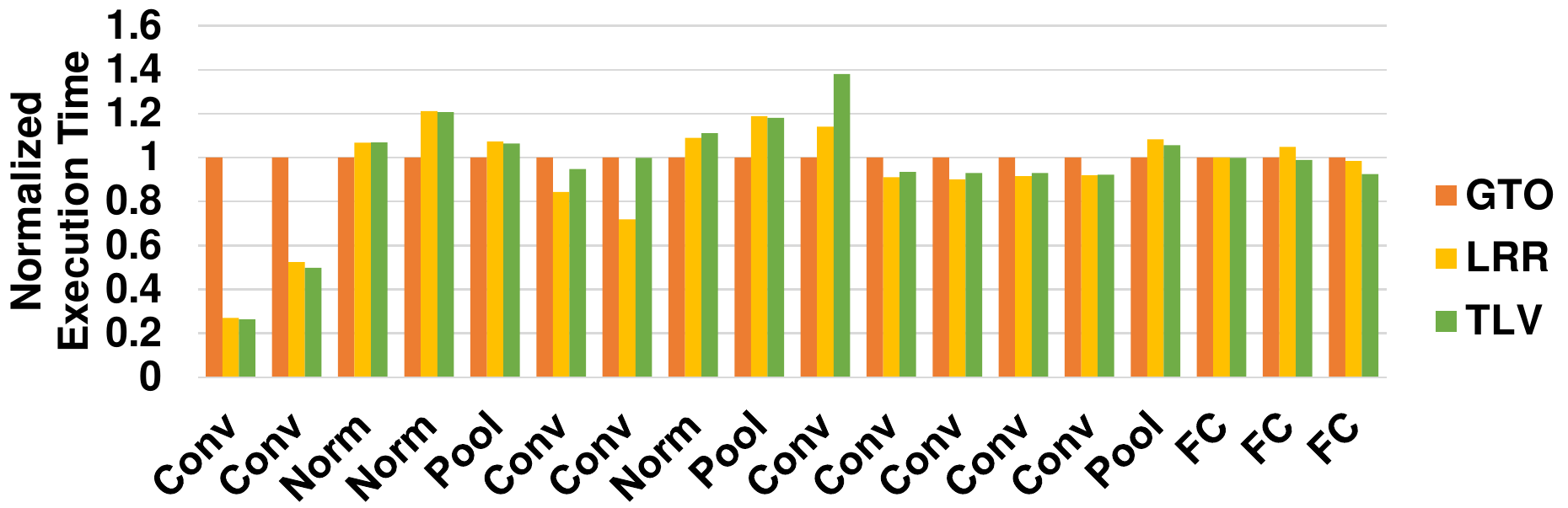}
\caption{Per-Layer Warp Scheduler Sensitivity of AlexNet} \label{sched.sens.alex}
\end{figure}
GPU application performance is highly influenced by warp schedulers. Thus, we evaluated the scheduler sensitivity on the performance of neural networks. Note that this evaluation is only possible with architecture simulators, which is one of the most important reasons that we propose a new benchmark suite that can run on micro-architecture simulators. Figure~\ref{sched.sens} shows the execution time when using GTO, LRR, and TLV schedulers, normalized by the execution time when using GTO. Due to the relatively short execution time, RNNs do not show a considerable performance difference across schedulers. AlexNet and ResNet show a significant performance improvement when using LRR. More specifically, Figure~\ref{sched.sens.alex} shows that the performance improvement is mainly acquired in convolution layers. We believe this is related with the data locality statistics shown in the previous subsection. Though convolution layers access memory extensively, due to the high data locality, the data is quickly fetched from on-chip caches. Thus, there is no need to move warps between ready and pending queues when they issue a memory operation as TLV and GTO. In such case, LRR can effectively reduce the queuing overhead while allowing sufficient time for each warp to wait for its data from memory. 

\textit{\underline{Observation 12.} The basic round-robin warp scheduler (LRR) is good enough for neural networks thanks to the high data locality of convolution layers.}
\section{Conclusion}
In this paper, we present a new DNN benchmark suite that can run without needing to install proprietary DNN libraries or heavy DNN frameworks. We also provide extensive architectural characteristics of five CNNs and two RNNs on both architecture simulator and real devices. The evaluations on an architecture simulator provides an in-depth insights of DNN accelerators design.





%

\bibliographystyle{IEEEtran}
\bibliography{bibliograph}

\end{document}